\documentclass[twocolumn,secnumarabic,amssymb, nonatbib, nobibnotes,aps,prd,showpacs,showkeys]{revtex4-1}

\usepackage{graphicx}
\usepackage{amsmath}
\newcommand{\beq}{\begin{equation}}
	\newcommand{\eeq}{\end{equation}}
	\newcommand{\eml}{\end{mathletters}}

\newcommand{\be}{\begin{equation}}
	\newcommand{\ee}{\end{equation}}
\newcommand{\bea}{\begin{eqnarray}}
	\newcommand{\eea}{\end{eqnarray}}

\newcommand{\eref}[1]{(\ref{#1})}

\begin{document}
	\title{Semi-microscopic theory of two proton emission}
	
	\author{S.A. Ghinescu $^{1,2}$ and D.S. Delion $^{1,2,3,4}$}
	\affiliation{
		$^1$ "Horia Hulubei" National Institute of Physics and Nuclear Engineering, \\
		30 Reactorului, POB MG-6, RO-077125, Bucharest-M\u agurele, Rom\^ania \\
		$^2$ Department of Physics, University of Bucharest,
		405 Atomi\c stilor, POB MG-11, RO-077125, Bucharest-M\u agurele, Rom\^ania\\
		$^3$ Academy of Romanian Scientists, 3 Ilfov RO-050044,
		Bucharest, Rom\^ania \\
		$^4$ Bioterra University, 81 G\^arlei RO-013724, Bucharest, Rom\^ania}
	\date{\today}
	
	\begin{abstract}
		We propose a semi-microscopic model for the simultaneous emission of two protons. This model has the advantage of avoiding certain technical aspects of a fully microscopic 3-body framework, while also allowing the investigation of the influence of proton pairing on the total lifetime of the decaying nucleus. Thus, we use the standard singlet two-proton wave function on the nuclear surface, provided by the Bardeen-Cooper-Schrieffer (BCS) approach, as a boundary condition for the propagator operator.
		Our model allows for the estimation of all quantities related to the $2p$ emission process, since it provides the 3-body wave function over most of the domain. We show that reasonable agreement with experimental values can be reached by varying the $pp$ pairing strength outside the nucleus in an interval close to the "bare" singlet value.
	\end{abstract}
	
	\maketitle
	\section{Introduction}
	The emission of two protons is an intriguing and exotic decay process, energetically possible in only a few nuclei close to the proton stability line. The first theoretical studies in this field were done in the sixties by Goldansky \cite{Goldansky_1960}, using a semi-classical 2-body formalism, who proposed two extreme mechanisms for the emission, sequential and simultaneous. Since then, various attempts have been made at describing the two proton emission. The simpler models, using semi-classical tools, have various degrees of success \cite{Sreeja_2019,Liu_HM_2021,Liu_YB_2021}. They generally depend on multiple parameters, but yield a relatively good predictive power. Nevertheless, when applied systematically to all known emitters, they reveal interesting patterns and help shed light on this difficult theoretical problem \cite{Delion_2022}. 
	
	The modern consensus is that regardless of the underlying mechanism, $2p$ emission is the three body process by nature and rigor dictates it must be treated in the hyperspherical harmonics framework \cite{Grigorenko_2001}. A number of interesting models have been developed in the last two decades, among others, coupled channels (CC)-like ones \cite{Grigorenko_2001,Grigorenko_2003,Grigorenko_2010,Wang_2018}, and R-matrix description \cite{Brown_2003}. While detailed and exhaustive, these models are also quite complex and various technical difficulties arise in contrast to 2-body processes. The spurious 2-body bound states in the nucleus-proton interaction have to be removed in an accurate fashion. The interaction between the emitted protons breaks the spherical symmetry and accurately finding 3-body resonances requires a large number of partial waves to be considered. However, such drawbacks are unavoidable in fully microscopic 3-body calculations.
	
	For the above reasons we propose in this paper an alternative to a fully microscopic theory. It has been shown by Grigorenko \cite{Grigorenko_2001} that neglecting all but point-like Coulomb interactions between the emitted fragments leads to a good order of magnitude estimate of the lifetime of the decaying nucleus. However, such a model does not provide much insight into various nuclear quantities. We propose that pairing correlations between the emitted protons can be investigated by allowing the protons to interact also via a nuclear potential. A great simplification can still be made by considering the protons interact with the nucleus only through Coulomb potentials. However, this clearly cannot hold when the protons are close to the daughter nucleus. Consequently, in this region we use the prescription of Delion et al \cite{Delion_2013}, with some modifications, to compute the wave function of paired protons in a resonant 2-body state. In this sense our model is \textit{semi-microscopic}. The total wave function cannot be rigorously determined together with its derivative. However, we will show that interesting studies can be made on the influence of the $pp$ paring strength on the partial life-time of the decaying nucleus. 
	
	The paper is structured as follows: in Section~\ref{sec:Formalism} we elaborate on the procedure to obtain the 3-body wave function and the decay width. We give recipes on the building of external and internal region wave functions, then use the current formulation to extract the (total and partial) decay width(s); in Section~\ref{sec:Results} we analyze various aspects of the $2p$ emission problem. We discuss the nature of the potential matrix, revealing that from a certain radius, the problem becomes practically uncoupled. We then study the partial waves obtained in the external region, showing that the asymptotic behavior is reached relatively soon outside the barrier. Then, we study the dependence and stability of the decay width on the proton pairing strength and matching point between internal and external regions, respectively.
	   
	\section{Formalism}
	\label{sec:Formalism}
	The simultaneous emission of two protons from a parent nucleus $P$ can be written schematically as
	\begin{align}
		_{Z}^{A}P\to _{Z-2}^{A-2}D(\mathbf{r}_3) + p(\mathbf{r}_1) + p(\mathbf{r}_2),
	\end{align}
	where $D$ is the daughter nucleus in ground state, and $\mathbf{r}_j$ denote the position of the three fragments in the laboratory frame. We will work in the approximation of infinitely heavy nuclei, hence $D$ is at rest and we set $\mathbf{r}_3 = \mathbf{0}$. Also, the total kinetic energy released in this process is called $Q$-value and is the sum of the kinetic energies of the two emitted protons
	\begin{equation}
		Q_{2p}=\epsilon_1 + \epsilon_2.
	\end{equation}
	This process is governed by the time-dependent Scr$\ddot{o}$dinger equation
	\begin{equation}
		i\hbar\frac{\partial \Psi(\mathbf{r},t)}{\partial t} = H\Psi(\mathbf{r},t),
		\label{eq:tdse}
	\end{equation} 
	where $H$ is the hamiltonian of the system and $\mathbf{r}$ denotes collectively all the position vectors involved in the system. 
	There are 2 equivalent ways of choosing the remaining coordinate frames. The $T$ system consists of the relative position vector of the emitted protons and the position of their center of mass w.r.t. the nucleus. The $Y$ system consists of the positions of the two protons relative to the nucleus. In this work we will employ the $Y$ system hence $\mathbf{r} \equiv \lbrace \mathbf{r}_1,\mathbf{r}_2 \rbrace$. This choice is more natural (as will become clear later) for the semi-microscopic description we propose.
	
	Since all known $2p$ emitters have a partial half life of the order of $10^{-3}$s or below, we can readily employ the Gamow approximation assuming the $2p$ emission state is a resonant one
	\begin{equation}
		\Psi(\mathbf{r}_1, \mathbf{r}_2, t) = e^{-\frac{i}{\hbar}(Q_{2p}-i\frac{\Gamma}{2})t}\psi(\mathbf{r}_1, \mathbf{r}_2),
		\label{eq:Gamow_approximation}
	\end{equation} 
	where, as usual, $\Gamma$ bears the significance of decay width and $\Gamma \ll Q_{2p}$. $\Gamma$ is real and positive in this approximation. We now replace Eq.~(\ref{eq:Gamow_approximation}) in Eq.~(\ref{eq:tdse}) and neglect for the moment $\Gamma$ and obtain the time-independent Schr$\ddot{o}$dinger equation for $\psi(\mathbf{r})$
	\begin{equation}
		H \psi(\mathbf{r}_1, \mathbf{r}_2) = Q_{2p}\psi(\mathbf{r}_1, \mathbf{r}_2).
		\label{eq:TIE}
	\end{equation} 
	The expression of the Hamiltonian is given by (recoil effects are neglected in the assumption of an infinitely heavy nucleus)
	\begin{equation}
		H = -\hbar^2 \sum_{j=1}^{2}\frac{1}{2m_{p}}\Delta_{j} + \sum_{j}^{2}V_{j}(\mathbf{r}_{j}) + v(\mathbf{r_1},\mathbf{r}_2),
		\label{eq:3BodyHamiltonian}
	\end{equation}
	where the sums runs over the 2 protons, $m_p$ is the proton mass, $\Delta_{j}$ denotes the usual 3-dimensional laplacian associated to the coordinate $\mathbf{r}_{j}$, $V_{j}(\mathbf{r}_j)$ is the interaction potential between the nucleus and proton $j$ and $v(\mathbf{r}_1, \mathbf{r}_2)$ is the interaction potential between the emitted protons. 
	
	The hyperspherical harmonics (HH) formalism (\cite{Avery_2018}) makes it possible to factorize the 6-dimensional space $(\mathbf{r}_1, \mathbf{r}_2)$ in one hyper-radial variable ($\rho$) and 5 hyper-angles. The usual convention one follows is
	\begin{align}
		\begin{aligned}
			\rho &= \sqrt{r_1^2 + r_2^2}, \hspace{1cm} \rho\in [0,\infty) \\
			\phi &= \arctan(r_2/r_1), \hspace{1cm} \phi \in \left[0, \frac{\pi}{2}\right]\\
			\theta_{1,2}, \varphi_{1,2} &= \text{Spherical angles of $\mathbf{r}_{j}$}.
		\end{aligned}
		\label{eq:HSCoords}
	\end{align}
	where $r_{j}=\left|\mathbf{r}_j\right|$. For briefness we denote $\Omega = (\phi, \theta_{1,2}, \varphi_{1,2})$. Using this transformation, Eq.~(\ref{eq:3BodyHamiltonian}) becomes
	\begin{align}
		\begin{aligned}
			H =& -\frac{\hbar^2}{2m_p}\left(\frac{\partial^2}{\partial \rho^2} + \frac{5}{\rho}\frac{\partial}{\partial \rho}\right) - \frac{\mathcal{L}^2(\Omega)}{\rho^2} +  \\
			&V_{1}(\rho\sin\phi) + V_{2}(\rho\cos\phi) +v(\rho,\Omega),
		\end{aligned}
		\label{eq:3BodyHamiltonianExplicit}
	\end{align}
	with $\mathcal{L}$, the grand-angular momentum, given by
	\begin{equation}
		\mathcal{L}^2(\Omega) = -\hbar^2\left[\frac{\partial^2}{\partial \phi} + 4\cot{2\phi} \frac{\partial}{\partial \phi} - \frac{1}{\hbar^2}\left(\frac{l_1^2}{\sin^2\phi} + \frac{l_2^2}{\cos^2\phi}\right)\right],
	\end{equation}
	and $l_{j}, j=1,2$ being the usual angular momenta of the two protons. The eigenvalue equation for $\mathcal{L}^2$ is
	\begin{equation}
		\mathcal{L}^2 \mathcal{Y}_c = \lambda_c \mathcal{Y}_c,
	\end{equation}
	with a multi-index $c$, $\lambda_c = K(K+4)$, $K = 2n+l_1 + l_2$ an even integer, $n$ an integer and 
	\begin{align}
		\begin{aligned}
			&\mathcal{Y}_c = \mathcal{N}_c (\sin\phi)^{l_1}(\cos\phi)^{l_2} P_{n}^{l_1+1/2,l_2+1/2}(\cos(2\phi)) \times \\
			&\left[\left(i^{l_1}Y_{l_1,m_{1}}\otimes i^{l_2}Y_{l_2,m_2}\right)_{J,M_J}\otimes \left(\chi_{1}\otimes\chi_{2}\right)_{S,M_S}\right]_{L,M},
		\end{aligned}
	\end{align}
	with $\chi_{j}, j=1,2$ being the spin and projection of the $j$-th proton and $\otimes$ denotes angular momentum coupling. Here, $L,M$ are the total recoupled angular momentum of the two protons and its projection respectively, $P^{\alpha,\beta}_n$ are the Jacobi polynomials and $N_{c}$ is a normalization constant. It is now clear that $c = \lbrace K,L,M,j,s,l_1,l_2\rbrace$. We also note here that we are working in the adiabatic approximation, hence no coupling appears between the inert daughter core and the emitted protons. 
	
	In this work we deal only with spin singlet states, i.e. $L=M_L=J=M_J=S=M_S=0$, hence $l_1 = l_2 \equiv l$. This greatly simplifies the formalism and the multi-index $c$ becomes now $c=\lbrace K,l \rbrace$ with $K=2n+2l$. 
	
	Similarly to the 3D case, we know expand the total spatial wave function as
	\begin{equation}
		\psi(\rho,\Omega) =  \rho^{-5/2}\sum_{c}g_{c}(\rho)\mathcal{Y}_c(\Omega),
		\label{eq:wf_expansion}
	\end{equation} 
	where we have included the factor before the sum to cancel the first derivative in the hamiltonian. Upon inserting this expansion, together with the factorized hamiltonian, into Eq.~(\ref{eq:TIE}) and projecting onto channel a specific channel $c$, we obtain a system of coupled equations
	\begin{equation}
		-\frac{\hbar^2}{2m_p}\left(\frac{d^2}{d\rho^2}-\frac{\lambda_c}{\rho^2}\right) g_{c}(\rho) +\sum_{c'}V_{c,c'}(\rho)g_{c}(\rho) = Eg_c(\rho),
		\label{eq:HHSystem}
	\end{equation}
	where 
	\begin{widetext}
	\begin{equation}
		V_{c,c'}(\rho) = \int d\Omega \mathcal{Y}_c(\Omega)\left[V_1(\rho\sin\phi) + V_2(\rho\cos\phi) + v(\rho,\Omega)\right]\mathcal{Y}_{c'}(\Omega),
		\label{eq:PME_Full}
	\end{equation}
	\end{widetext}
	are the potential matrix elements (PME). As usual, instead of solving Eqs.~\ref{eq:HHSystem}, we will solve the system for the associated fundamental matrix, which has on columns linearly independent solutions of \ref{eq:HHSystem}. By straight-forward generalization this system is given by
	\begin{eqnarray}
		&&-\frac{\hbar^2}{2m_p}\left(\frac{d^2}{d\rho^2}-\frac{\lambda_c}{\rho^2}\right) g_{c,c'}(\rho) 
\\ \nonumber &+&\sum_{c''}V_{c,c''}(\rho)g_{c'',c'}(\rho) = Q_{2p}g_{c,c'}(\rho).
		\label{eq:HHSystemFundamMat}
	\end{eqnarray}
	It is instructive to rewrite $\lambda_c$ as $\lambda_c = l_c(l_c+1)$, with $l_c = K + 3/2$. This reveals one of the ways in which the 3 body decay is fundamentally different from the any 2 body process. The centrifugal barrier is present even in the lowest channel (i.e., when $K=0$).
	
	In our model we keep all partial waves with $l\le 7$ and $K \le 30$. We found that increasing $K$ above this value, while keeping $l$ constant induces no change in the decay width.
	
	For the computation of the wave function, we define two regions: internal and external, meeting at a mathcing radius $R_m$ and discuss them separately.

	\subsection{External wave function ($\rho>R_m$)}
	In this section we apply the tools presented above to build a three body wave-function in the external region for the two proton emission process. 
	
	In order to solve the system given in Eq.~(\ref{eq:HHSystemFundamMat}), we need to specify the potentials. It is well known employing 2 body potentials that allow bound 2 body states introduces spurious effects. Such bound states are usually eliminated either through projections \cite{Sparenberg_1997} or through supersymmetric transformations \cite{Sparenberg_1997}. In order to avoid this extra difficulty, we consider  the nucleus-proton potential to be that of point charge interacting with a charged sphere of radius equal to the nuclear radius. We considered the proton-proton potential to be given by a simple central gaussian. We denote $|\mathbf{r}_1 - \mathbf{r}_2|\equiv r_{12}$ and
	\begin{equation}
		\label{vpp}
		v(|\mathbf{r}_1-\mathbf{r}_2|) = v(r_{12})=v_0e^{-\left(r_{12}/r_0\right)^2} + \frac{e^2}{r_{12}},
	\end{equation}
	where $v_0$ is a negative constant and $r_0 = 2$fm is the proton-proton interaction radius. Even though this potential allows shallow bound states (for $v_{0} \leq -30$ MeV and $r_0 = 2.0$ fm), we found no influence of this potential on our calculations. 
	
	The system (\ref{eq:HHSystemFundamMat}) can be solved by the usual Numerov method. However, we found that much better stability can be achieved by employing the renormalized Numerov algorithm \cite{Johnson_1978}. In both cases the solution can be found by specifying the boundary conditions (BC) at $\rho\to\infty$ for each $g_{c,c'}$. This in itself can be achieved only in some approximation, since the system is not asymptotically decoupled (in principle). Grigorenko et al \cite{Grigorenko_2001} have analyzed the effects on accuracy and precision of multiple boundary conditions and determined that a good precision can be obtained through "diagonalized Coulomb" potentials. We follow this approach and, for the farthest two radial points, diagonalize the potential matrix. The Sommerfeld parameter for each channel $\eta_c$ is then used to for building the BC as outgoing Coulomb-Hankel functions \cite{NIST:DLMF}
	\begin{equation}
		g_{c,c'}(\rho)|_{\rho\to\infty} = \mathcal{H}_{\l_{c}}(\eta_c, k\rho)\delta_{c,c'}, \hspace{0.5cm} k=\sqrt{2m_pE}/\hbar.
	\end{equation}
	
	\subsection{Internal wave function ($\rho<R_{m}$)}
	Here we present the procedure to obtain the internal 3-body wave function, when the two protons are emitted from a paired state. In order to avoid using the complicated 3-body framework presented above in the very complex internal region, we first compute the energy levels and wave functions of bound and resonant protons inside the parent nucleus in the 2-body framework. Details of the procedure can be found in \cite{Delion_2010} together with the Woods-Saxon(WS) parametrization including spin-orbit interaction. Here we briefly outline the procedure. 

	The single particle (sp) state diagonalizing a spherical WS + spin-orbit + Coulomb mean field with eigenvalue $\epsilon$ is a superposition of harmonic oscillator (ho) orbitals
	\begin{equation}
		| \psi_{\epsilon, l, j, m}\rangle = c^{\dagger}_{\epsilon, l, j, m} |0\rangle = \sum_{n}d_{\epsilon, l, j}^{n} |\varphi^{\beta}_{n,j,j,m}\rangle,
	\end{equation}
	depending upon the radial quantum number $n$, angular momentum $l$, total spin $j$, spin projection $m$ and the ho parameter
	\begin{equation}
		\beta = \frac{m_p \omega}{\hbar}.
	\end{equation}
	The coordinate representation of the ho wave function depending on $x\equiv(\mathbf{r}, s)$ is given by
	\begin{align}
		\begin{aligned}
			\varphi^{\beta}_{n,l,j,m}(x) &= \langle x| \varphi^{\beta}_{n,l,j,m} \rangle= \left[\phi_{n,l}^{\beta}(\mathbf{r})\otimes \chi_{\frac{1}{2}}(s)\right]_{j,m}\\
			\phi_{n,l,m}^{\beta} &= \mathcal{R}_{n,l}^{\beta}i^{l}Y_{l,m}(\hat{r}),
		\end{aligned}
		\label{eq:SPWF}
	\end{align}
	in terms of the radial ho function $\mathcal{R}_{n,l}^{(\beta)}(r)$, given by the Laguerre polynomial and the Euler Gamma function
	\begin{equation}
		{\mathcal R}_{n,l}^{(\beta)}(r)=(-)^{n}\left[\frac{2 \beta^{3/2}n!}{\Gamma(n+l+3/2)}\right]^{1/2}r^{l+1}e^{-\beta r^2} L_{n}^{l+1/2}(\beta r^2).
	\end{equation}

	Thus, the WS wave function can be rewritten as
	\begin{equation}
		\psi_{\epsilon, l, j, m}(x) = \langle x|\psi_{\epsilon, l, j, m} \rangle = \mathcal{R}_{\epsilon, l,j}(r)\mathcal{Y}_{j,m}^{(l,\frac{1}{2})}(\hat{r}, s),
	\end{equation}
	in terms of the radial WS wave function and spin-orbit harmonics, respectively
	\begin{align}
		\begin{aligned}
			\mathcal{R}_{\epsilon,l,j}(r) &= \sum_{n}d_{\epsilon, l,j}^{n} \mathcal{R}_{nl}^{\beta}(r)\\
			\mathcal{Y}_{j,m}^{(l,\frac{1}{2})}(\hat{r}, s) &= \left[i^lY_{l}(\hat{r})\otimes\chi_{\frac{1}{2}}\right]_{j,m}.
		\end{aligned}
	\end{align}

	A normalized pair state with a given spin/parity, labeled by $J$, is defined as follows
	\begin{align}
		\begin{aligned}
			|ab;JM\rangle &= \mathcal{N}_{ab}(J)\left[c^{\dagger}_{a}\otimes c^{\dagger}_{b}\right]_{J,M}|0\rangle\\
			\mathcal{N}_{a,b}(J) &\equiv \frac{\sqrt{1-\delta_{a,b}(-)^{J}}}{1+\delta_{a,b}},
		\end{aligned}
	\end{align}
	where $a\equiv(\epsilon_a, l_a,j_a), b=(\epsilon_b, l_b,j_b)$. In the configuration space, the pair state is given by
	\begin{widetext}
	\begin{align}
		\begin{aligned}
			\Psi_{abJM}(x_1,x_2) &= \langle x_1x_2|ab;JM\rangle  \mathcal{N}_{ab}(J)\mathcal{A}\lbrace\left[\psi_a(x_1)\otimes\psi_b(x_2)\right]_{JM}\rbrace\\
			&=\mathcal{N}_{ab}(J)\frac{1}{\sqrt{2}}\lbrace \left[\psi_a(x_1)\otimes\psi_b(x_2)\right]_{JM} - \left[\psi_a(x_2)\otimes\psi_b(x_1)\right]_{JM} \rbrace.
		\end{aligned}
	\end{align}
The complete wave function is given 
	\begin{align}
		\begin{aligned}
			X_{\epsilon,l,j}\Psi_{abJM}(x_1,x_2) &= X_{\epsilon,l,j}\sqrt{2}\mathcal{N}_{ab}(J)\sum_{n_a,n_b} d^{n_a}_a d^{n_b}_b\Phi_{abJM}(x_1,x_2)\\
			\Phi_{abJM}(x_1,x_2) &= \left[\left(\phi^{(\beta)}_{n_a,l_a}(\mathbf{r}_1)\otimes \chi_{\frac{1}{2}}(s_1)\right)_{j_a}\otimes\left(\phi^{(\beta)}_{n_b,l_b}(\mathbf{r}_2)\otimes \chi_{\frac{1}{2}}(s_2)\right)_{j_b}\right]_{J,M},
		\end{aligned}
	\end{align}
	\end{widetext}
in terms the pair formation amplitude
\begin{eqnarray}
X_{\epsilon,l,j}&=&\frac{1}{2}\langle BCS|[c^{\dagger}_{\epsilon,l,j}\otimes c^{\dagger}_{\epsilon,l,j}]_0|BCS\rangle
\\ \nonumber&=&
\frac{\sqrt{2j+1}}{2}u_{\epsilon, l, j}v_{\epsilon,l,j},
\end{eqnarray}
depending on standard BCS amplitudes.
By changing $\Phi$ from $jj$ to $LS$ coupling one considers the singlet component
\begin{eqnarray}
&&\Phi_{abJM}(x_1,x_2)\rightarrow
\\ \nonumber
&&\left[\phi_{n_a,l_a}^{\beta}(\mathbf{r_1}) \otimes \phi_{n_b,l_b}^{\beta}(\mathbf{r_2}) \right]_{J} \otimes \left[\chi_{\frac{1}{2}}(s_1)
\otimes\chi_{\frac{1}{2}}(s_2)\right]_0
\\ \nonumber
&\times& \left\langle (l_al_b)J\left(\frac{1}{2}\frac{1}{2}\right)0;J| \left(l_a, \frac{1}{2}\right) j_a
\left(l_b\frac{1}{2}\right)j_b;J\right\rangle.
\end{eqnarray}
	Notice that all pair phases $(i)^{l_a+l_b} = (-)^{(l_a+l_b)/2}$ of products between sp wave functions (\ref{eq:SPWF}) have the same sign, due to the common angular momenta parities and therefore the product of these terms in the matrix element is positive. 

	Next, we change the radial part by using the Talmi-Moshinsky (TM) transformation
\begin{eqnarray}
&&\left[\phi^{\beta}_{n_a,l_a}(\mathbf{r}_1)\otimes\phi^{\beta}_{n_b,l_b}(\mathbf{r}_2)\right]_{J,M} 
\\ \nonumber
&=& \sum_{nlNL}\left[\phi_{n,l}^{\beta/2}(\mathbf{r})\otimes\phi_{N,L}^{2\beta}(\mathbf{R})\right]_{JM}\langle nlNL;J|n_al_an_bl_b;J\rangle,
\end{eqnarray}
	from absolute to relative and center of mass (cm) coordinates
	\begin{align}
		\begin{aligned}
			\mathbf{r} &= \mathbf{r}_1-\mathbf{r_2}\\
			\mathbf{R} &= \frac{\mathbf{r}_1 + \mathbf{r}_2}{2},
		\end{aligned}
	\end{align}
	by using in summation the conserving energy conditions
	\begin{equation}
		2n_a+l_a+2n_b+l_b = 2n+l+2N+L.
	\end{equation}
	The paired state wave function now has to be expanded in hyperspherical harmonics. We do this by the usual Fourier decomposition
	\begin{align}
		\Psi_{abJM}(x_1,x_2) &= \rho^{-5/2}\sum_{c}f_{c}(\rho)\mathcal{Y}_{c}(\Omega)\\
		f_c(\rho) &= \rho^{5/2}\int_{\Omega} d\Omega \Psi_{abJM}(x_1,x_2)\mathcal{Y}_{c}(\Omega).
	\end{align}
	\subsection{Decay width computation}
	Suppose we have complete knowledge of the spatial component of the wave function in Eq.~(\ref{eq:Gamow_approximation}). Without neglecting $\Gamma$, we can replace Eq.~(\ref{eq:Gamow_approximation}) into Eq.~(\ref{eq:tdse}), making use of Eq. (\ref{eq:3BodyHamiltonianExplicit}) but with an arbitrary potential $V(r,\Omega)$. For brevity we use $\Delta = \Delta_1 + \Delta_2$, the 6D laplacian and obtain for the wave function and its conjugate:
	\begin{align}
		\left(Q-i\frac{\Gamma}{2}\right) \psi =& \left[-\frac{\hbar^2}{2m_p}\Delta + V(\rho,\Omega)\right]\psi\\
		\left(Q+i\frac{\Gamma}{2}\right) \psi^{\dagger} =& \left[-\frac{\hbar^2}{2m_p}\Delta + V(\rho,\Omega)\right]\psi^{\dagger}.
	\end{align}
	Multiplying to the left both equations, the first by $\psi^{\dagger}$ and the second by $\psi$ and subtracting the first from the second, we obtain
	\begin{equation}
		\Gamma |\psi|^{2} = \frac{\hbar^2}{2m_pi} \left(\psi\Delta\psi^{\dagger} - \psi^{\dagger}\Delta\psi\right).
	\end{equation}
	Next, we integrate over the volume of a hypersphere of radius $R$, large enough to contain most of the wave function ($\int dV |\psi|^2 = 1$). The definition becomes then
	\begin{equation}
		\Gamma = \frac{\hbar^2}{2m_pi}\int_{0}^{R}d\rho \rho^{5}\int_{\Omega} d\Omega (\psi\Delta\psi^{\dagger} - \psi^{\dagger}\Delta\psi).
	\end{equation}
	We now use the partial wave expansion of Eq.~(\ref{eq:wf_expansion}) and the orthonormality of the HH to write
	\begin{equation}
		\Gamma = \sum_{c}\frac{\hbar^2}{2m_pi}\left(g_c(R)\frac{dg_c^{\dagger}(\rho)}{d\rho}|_{R} - g_c^{\dagger}(R)\frac{dg_c(\rho)}{d\rho}|_{R}\right).
		\label{eq:GammaDefintion}
	\end{equation}
	From the expression above, it would appear that the decay width depends on the hyper-radius of computation. However, if this hyper-radius is large enough, $g_c$ are approximately the Coulomb-Hankel functions. Since the quantity in brackets is nothing else than the Wronskian, it follows that the decay width is independent of computation point, at large hyper-radii. We have observed that convergence is achieved at moderate distances (30fm) for interesting nuclei.
	
	The advantage of using a semi-microscopic theory is that finding the entire wave function is not mandatory. Instead, we integrate the system ~\eref{eq:HHSystemFundamMat} from far away and match a linear combination of the matrix $g$ to the internal wave function $f$ at $R_m$. 
	\begin{equation}
		f_{c}(R_m) = \sum_{c'} g_{c,c'}(R_m)N_{c'}.
	\end{equation}
	$N_{c}$ are called scattering amplitudes and the decay width is directly related to them, as we will show promptly. The obvious drawback of such a method is that only the wave function will be continuous, while its derivative will not. However, this is important only if the evaluation of the probability current is needed close to the nucleus. Describing emission from narrow resonances does not carry such constraints. The internal wave-function is real and normalized to 1 up to the hyper-radius $R_{m} \le R$. In this case, Eq.~(\ref{eq:GammaDefintion}) holds, but with $g_{c}(R) = \sum_{c} g_{c,c'}(R)N_{c'}$. The usefulness of the scattering amplitudes becomes even more obvious in the limit $R\to\infty$, where $g_{c,c'}(R)\to N_c\mathcal{H}_{l_{c}}(\eta_c, kR)\delta_{c,c'}$, the Wronskian is $2ki$ and the decay width becomes
	\begin{equation}
		\Gamma = \frac{\hbar^2 k}{m_{p}}\sum_{c}|N_{c}|^2.
		\label{eq:GammaFromAmplitudes}
	\end{equation}
	Following this recipe, we verified that Eq.~(\ref{eq:GammaFromAmplitudes}) and Eq.~(\ref{eq:GammaDefintion}) give the same result within machine precision for $R\ge 30$ fm. However, in contrast with a fully microscopic theory, $\Gamma$ depends in the matching radius between the internal and external wave functions. This is a well-known drawback of semi-microscopic theories. Nevertheless, it avoids the need to remove 2-body bound states from the nucleus-proton potentials and allows us the study of proton pairing on the life-time. Anyway, we will show that the decay width weakly
depends upon the matching radius in a relative large interval beyond the nuclear radius.
	
	\section{Results and Discussions}
	\label{sec:Results}
	In this section we will analyze various aspects of the 3-body problem using the semi-microscopic model built above. To this purpose we consider 3 nuclei, with input parameters specified in Table I.
	
	\begin{table}
		\caption{\label{tab:Nuclei} Parameters of analyzed nuclei. The second column contains the atomic number of the daughter nucleus. The third column contains the $Q$-value of the $2p$ decay. The forth column gives the angular momentum of the state from which protons are emitted. The last column contains the logarithm of the experimental decay width.}
		\begin{tabular}{|c|c|c|c|c|c|}
			\hline
			Nucleus & $Z_D$ & $Q_{2p}$(MeV) & l & $\log_{10}\Gamma_{\mathrm{exp}}$(MeV)  & Ref. \\
			\hline
			$^{19}$Mg & 10 & 0.750 & 2 & -10.121 & \cite{Goncalves_2017}\\
			$^{45}$Fe & 24 & 1.210 & 3 & -18.941 & \cite{Miernik_2007}\\
			$^{54}$Zn & 28 & 1.480 & 1 & -18.911 & \cite{Blank_2005}\\
			\hline
		\end{tabular}
	\end{table}
We first diagonalized the WS mean field for protons by adjusting its real part in order to obtain at the Fermi level the positive experimental proton energy $\epsilon=Q_{2p}/2$ (the paired nucleons have equal energies).
Then we solved BCS equations by using the inter-proton force given by the nuclear gaussian interaction in Eq. (\ref{vpp}). 
By changing the nuclear strength $v_0$, we obtained the pairing gap at the Fermi level equal to the experimental pairing gap 
$\Delta_F=\Delta_{exp}=12/\sqrt{A}$, considered as an input parameter.

	\subsection{Potential Matrix}
	We discuss here the nature of the potential matrix given by Eq.~(\ref{eq:PME_Full}). One fundamental difference between 3-body scattering and the 2-body analog is the channel coupling even at large hyperradius. Indeed the exact solution at large distances should account for situations in which there can be residual 2-body interactions. It is our purpose in this paper, however, to establish a set of approximations that simplify the picture as much as possible while retaining most of the mathematical rigor. The first aspect we draw attention to is the diagonality of the potential matrix.

\begin{figure}
		\includegraphics[width=\columnwidth]{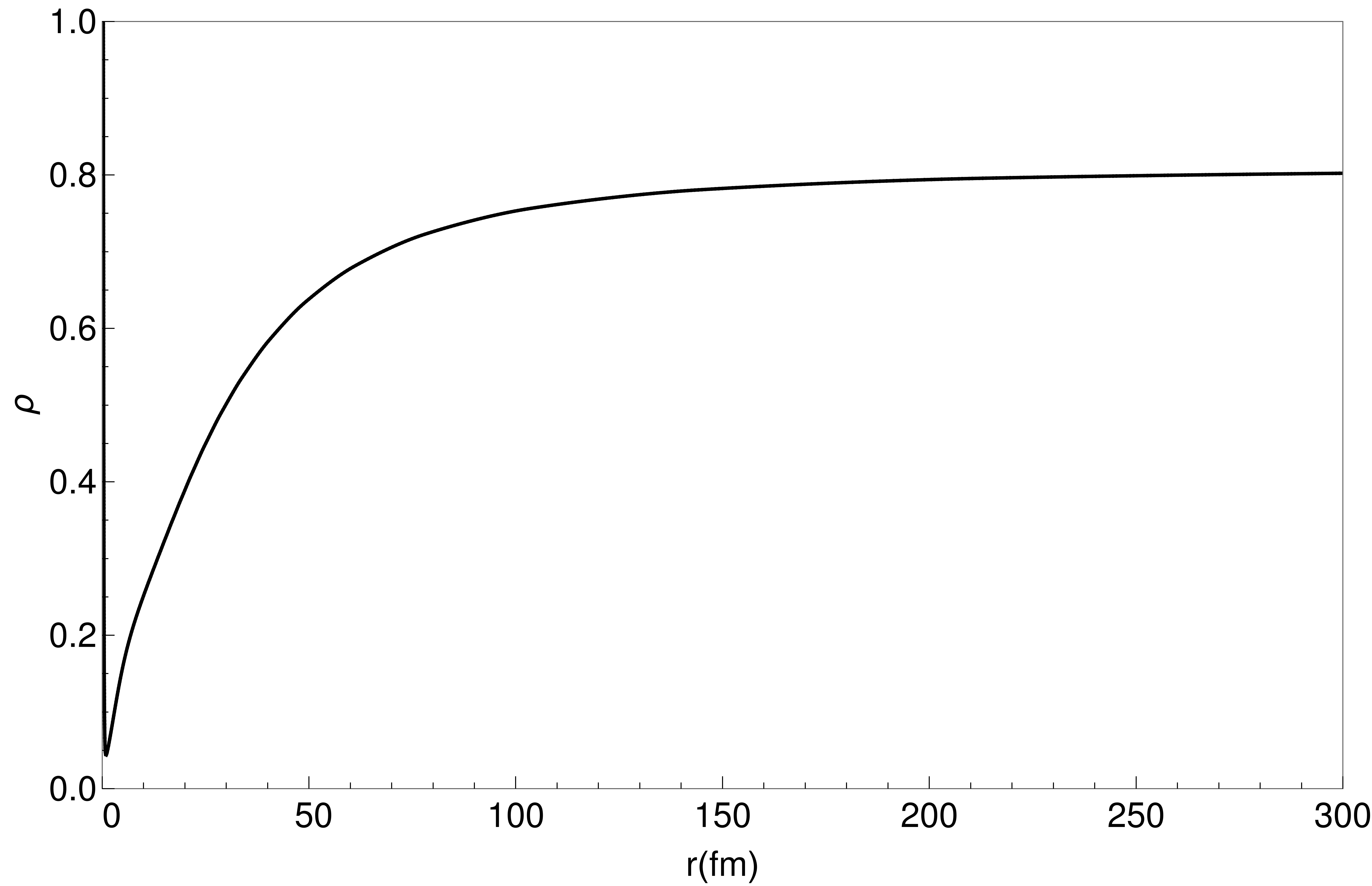}
		\caption{\label{fig:PM_correlations}Pearson's correlation coefficient for the potential matrix as function of the hyper-radius (for $^{45}$Fe with $v_0$=-35 MeV and $r_0$=2 fm)}
	\end{figure}

\begin{figure}
		\includegraphics[width=\columnwidth]{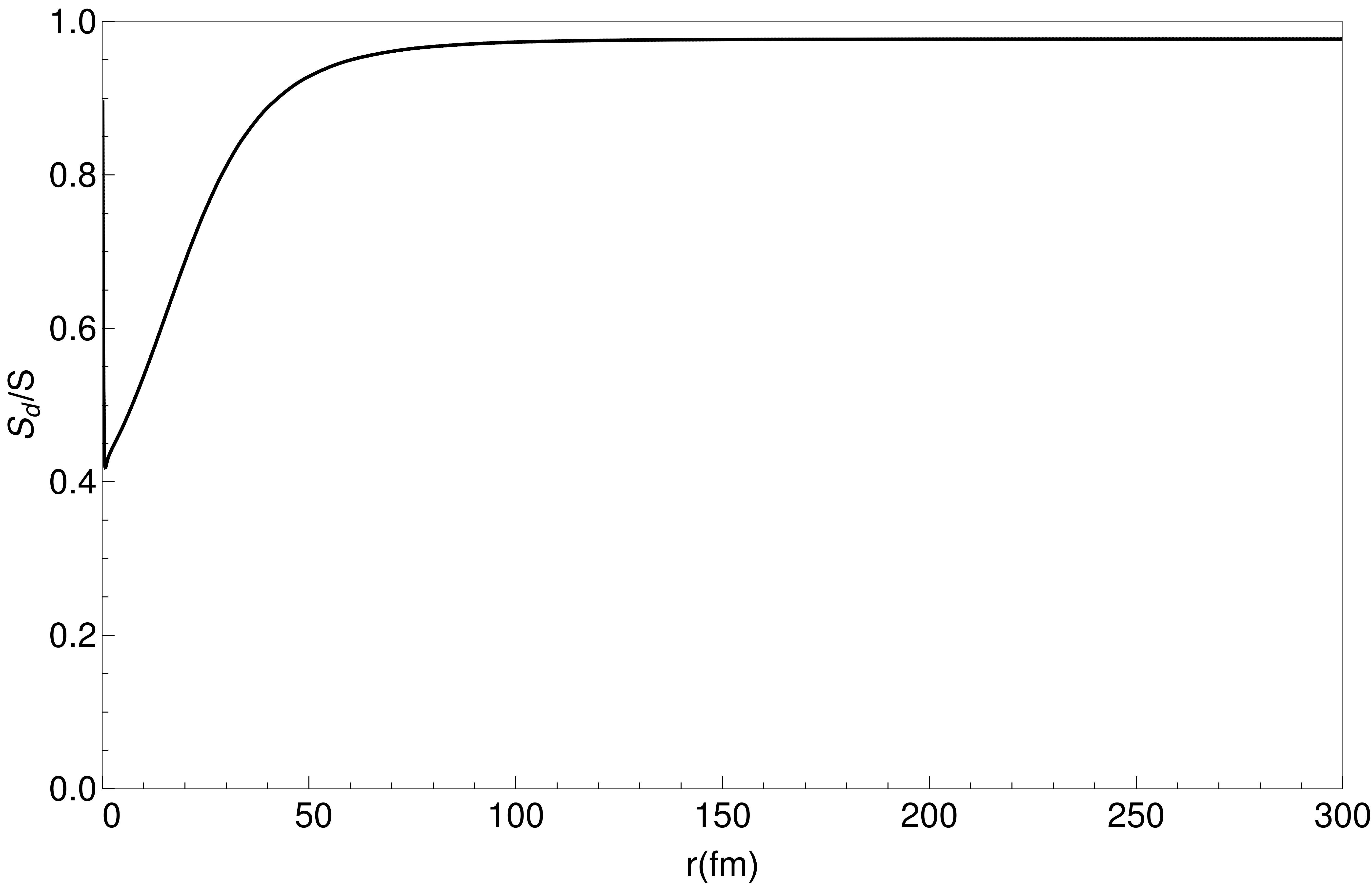}
		\caption{\label{fig:PM_squaresSum} Potential matrix diagonal dominance as function of hyper-radius (for $^{45}$Fe with $v_0$=-35 MeV and $r_0$=2 fm). See text for definitions.}
	\end{figure}

	One way of measuring \textit{how diagonal the matrix is} consists in the estimate of the Pearson's correlation coefficient between rows and columns \cite{Stigler_1989}. In Fig.~\ref{fig:PM_correlations} we plot this quantity of the matrix (\ref{eq:PME_Full}) for $^{45}$Fe. In case of a diagonal matrix, this coefficient is 1. In our case, after the monopole turning point ($\simeq$100fm) stability at about 0.8. This implies the matrix is diagonally dominant, which, to first order allows for a decoupled treatment at infinity.
	The above consideration is reinforced in Fig.~\ref{fig:PM_squaresSum} where we plot the ratio of 
	\begin{equation*}
		S_d = \sqrt{\sum_{c}V_{cc}^2} \hspace{1cm} \text{and} \hspace{1cm}
		S = \sqrt{\sum_{c,c'}V_{cc'}^2},
	\end{equation*}
	showing that the diagonal accounts for $95\%$ of the Froebinius norm. Again this points towards the possibility of using decoupled solutions at large radii.	
	
\begin{figure}
		\includegraphics[width=\columnwidth]{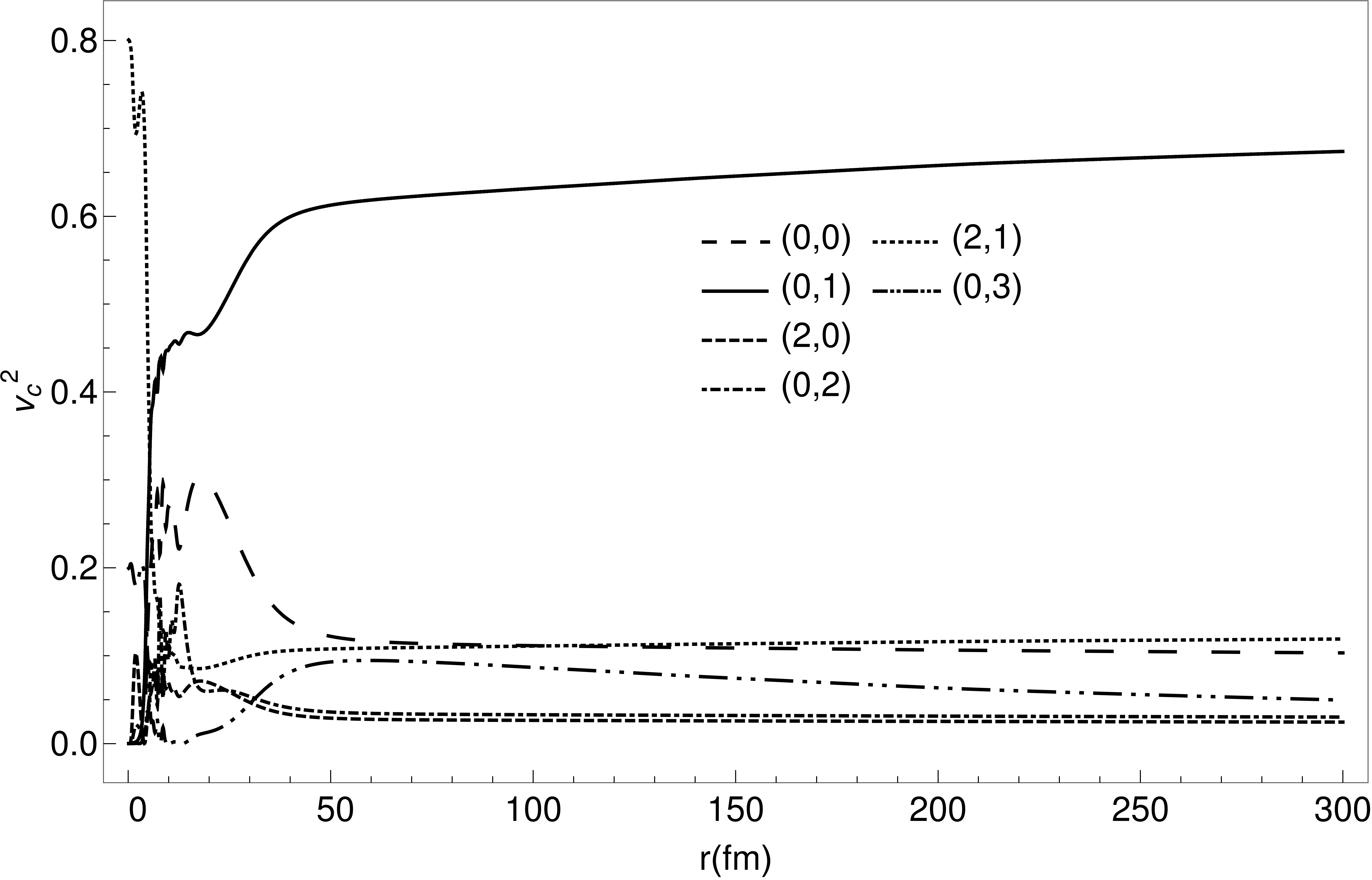}
		\includegraphics[width=\columnwidth]{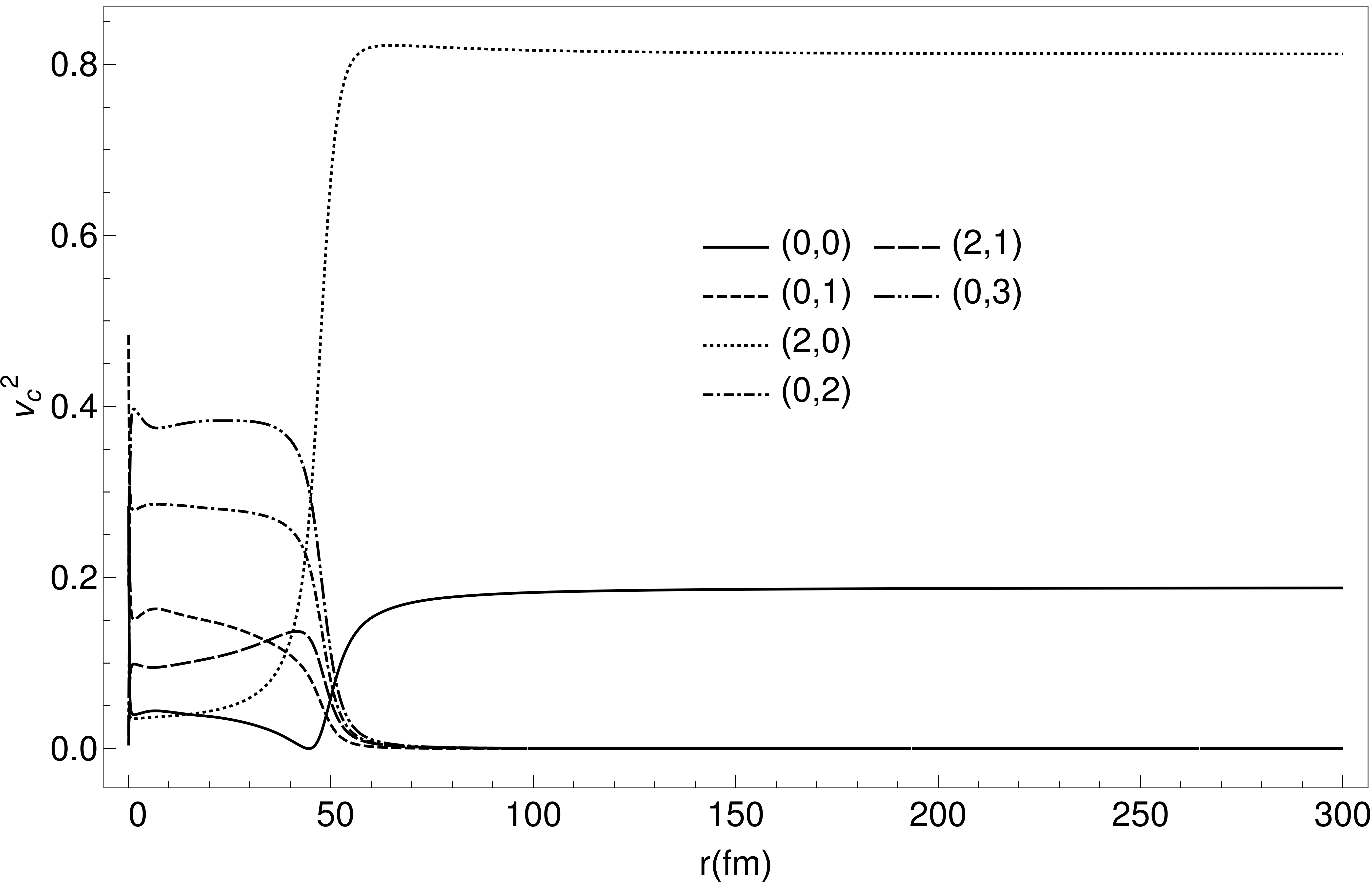}
		\caption{\label{fig:PM_eigenvectors} Weight of each $(n,l)$ channel in the eigenvectors associated to the lowest (upper panel) and highest (lower panel) eigenvalues. The same parameters as in Fig.~\ref{fig:PM_correlations} are used.}
	\end{figure}

\begin{figure}
		\includegraphics[width=\columnwidth]{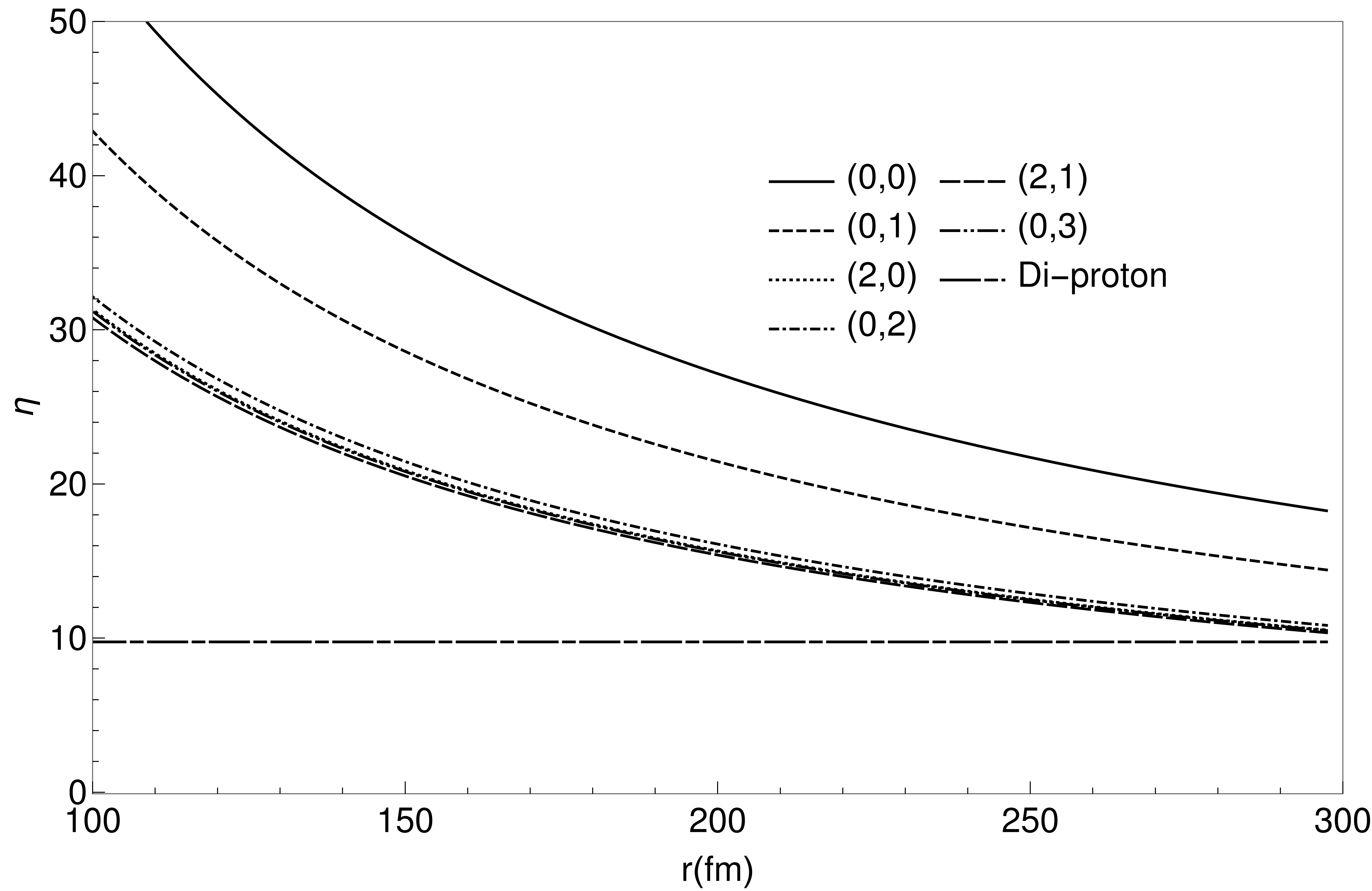}
		\caption{\label{fig:PM_eigenvalues} Sommerfeld parameters as function of the hyper-radius. The legend specifies the $(n,l)$ channel to which the eigenvalue has been ascribed. The same parameters as in Fig.~\ref{fig:PM_correlations} are used.}
	\end{figure}

	We now investigate the eigenvalues of the potential matrix. To this purpose, we diagonalize the matrix through $VD=D\lambda^{(v)}$ where $D$ is the matrix of eigenvectors and $\lambda^{(v)}$ is a diagonal matrix with eigenvalues as non-zero elements. Assuming, then, the potential to be of type $V(r)/Q\simeq2\eta_c/kr$ when $r\to \infty$, we can extract the sommerfeld parameters as
	\begin{equation}
		\eta_c = \frac{1}{2}kr\frac{V(r)}{Q_{2p}} = \frac{1}{2}k r \frac{\lambda^{(v)}_c}{Q_{2p}}.
		\label{eq:Sommerfeld_diag}
	\end{equation} 
	The problem that remains is how to assign these eigenvalues to the $(n,l)$ channels. To this purpose, in Fig. \ref{fig:PM_eigenvectors} we show the squared amplitude (weight) of each channel in the eigenvectors corresponding to the minimum and maximum eigenvalue of the potential. While there is strong mixing in the sub-barrier region, at large distances each eigenvector has a dominant component in one channel. This allows us to assign each eigenvalue, hence $\eta_c$, to one channel as shown in Fig. \ref{fig:PM_eigenvalues}.
	We note here that the largest Sommerfeld parameters of Eq.~(\ref{eq:Sommerfeld_diag}) do not reach an asymptotic behavior. This is expected since the problem is essentially coupled even at large distances. Nevertheless, outside the turning point, the Sommerfeld parameter has little relevance since the Coulomb function modulus is of the order of unity.

	\subsection{Wave function}
\begin{figure}
		\includegraphics[width=\columnwidth]{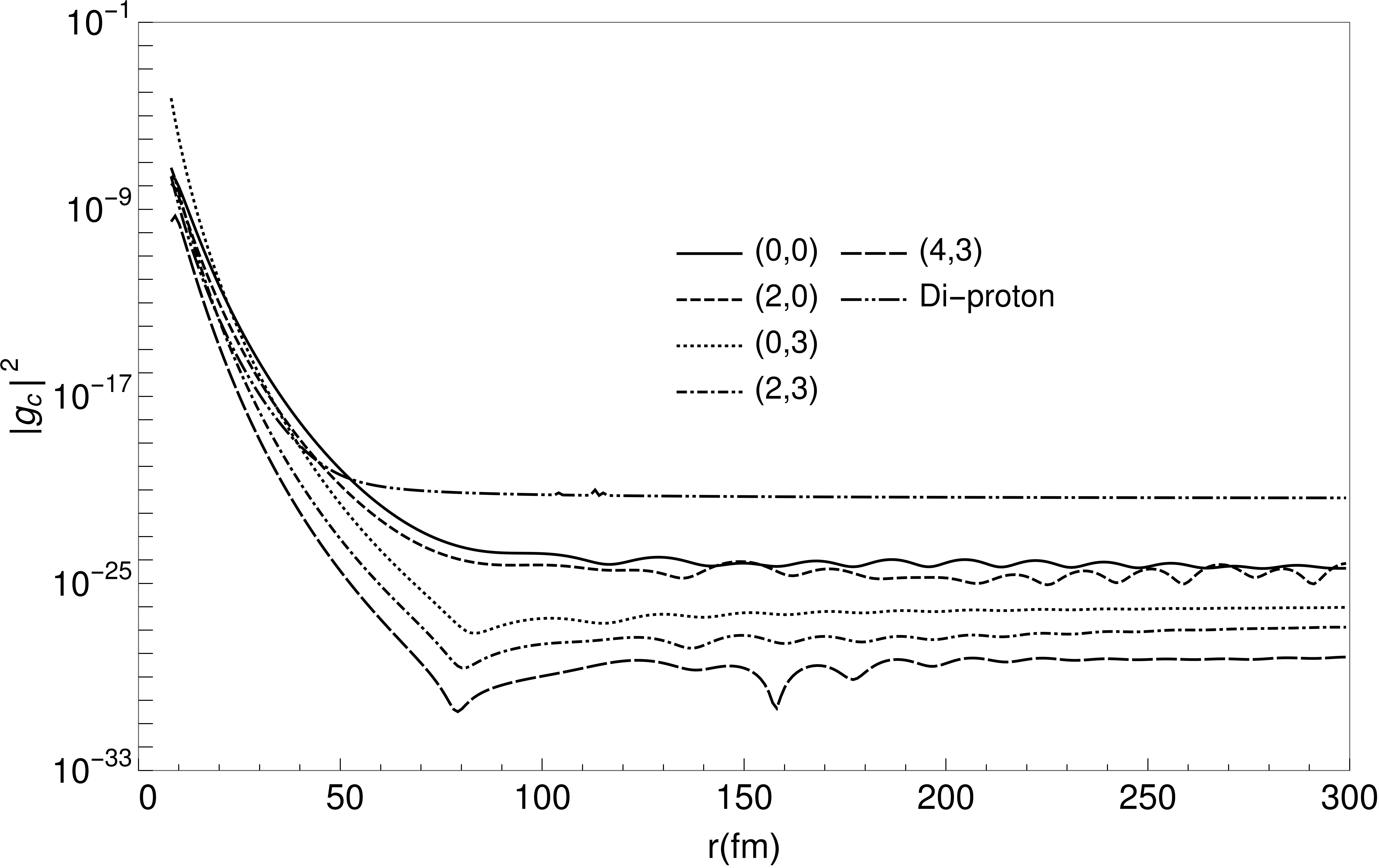}
		\caption{\label{fig:WF_channels} Wave function components squared for $l=0$ and $l=3$ for the $^{45}$Fe nucleus with $v_0=-35$~MeV and $r_0=2$~fm.}
	\end{figure}

	Some insight into the nature of our problem can be gained through the examination of the wave function behavior channel by channel. Moreover, the case of $^{45}$Fe allows the study of the channel mixing since the paired protons have $l$=3 at the Fermi level, hence they cannot be on the lowest hyper-spherical channel on the nuclear surface. In Fig.~\ref{fig:WF_channels}, we plot external wave function components after matching at $r=$7fm. We see that immediately after the nuclear surface, the components populated by the BCS function are dominant. However, after a few tens of fm, the entire wave function \textit{flows} essentially in the lowest 2 hyper-spherical channel. 
	This effect can be understood in terms of both the centrifugal barrier, which is significantly smaller in the $(0,0)$ and $(2,0)$ channels and of the potential matrix which couples every channel to the monopole. This is again quite different from the axially deformed 2-body case, where only neighboring channels are coupled. An important consequence is the following: despite the apparent complexity of the 3-body problem, the strong coupling of various channels to the monopole allows for a great simplification by discarding all but a few low-lying channels.

	\subsection{Decay width}
\begin{figure}[ht!]
		\includegraphics[width=\columnwidth]{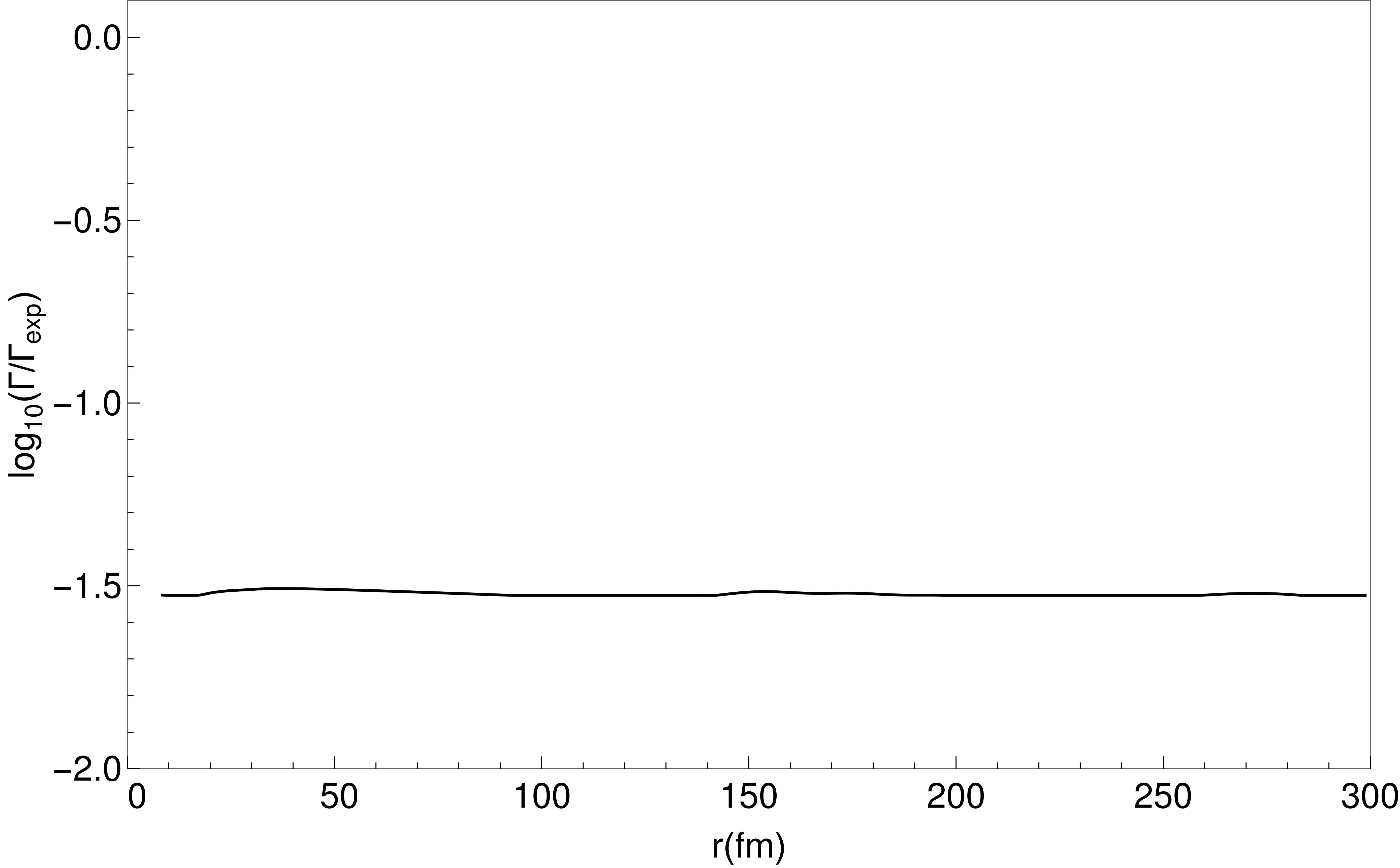}
		\caption{\label{fig:Gamma_vs_R}Decay width as function of the hyper-sphere radius $R$ in Eq.~(\ref{eq:GammaDefintion}) computed using the same parameters as in Fig.~\ref{fig:WF_channels}.}
	\end{figure}

	A final check for the stability of our method concerns the decay width. The variation with distance of the decay width is given by Eq.~(\ref{eq:GammaDefintion}). The term in brackets is the Wronskian $W(g_c\dagger,g_c)$. In the 2-body case, it is proportional to the Wronskian of outgoing Coulomb functions, which, in turn is constant and equal to $2i$. We show in Fig.~\ref{fig:Gamma_vs_R} that the same holds for the 3-body case, still under the barrier, but at greater distances than in the 2-body case. In other words, the 3-body decay resembles the 2-body decay after a certain hyper-radius, when all couplings can be ignored. 

	\begin{figure}[ht]
		\centering
		\includegraphics[width=\columnwidth]{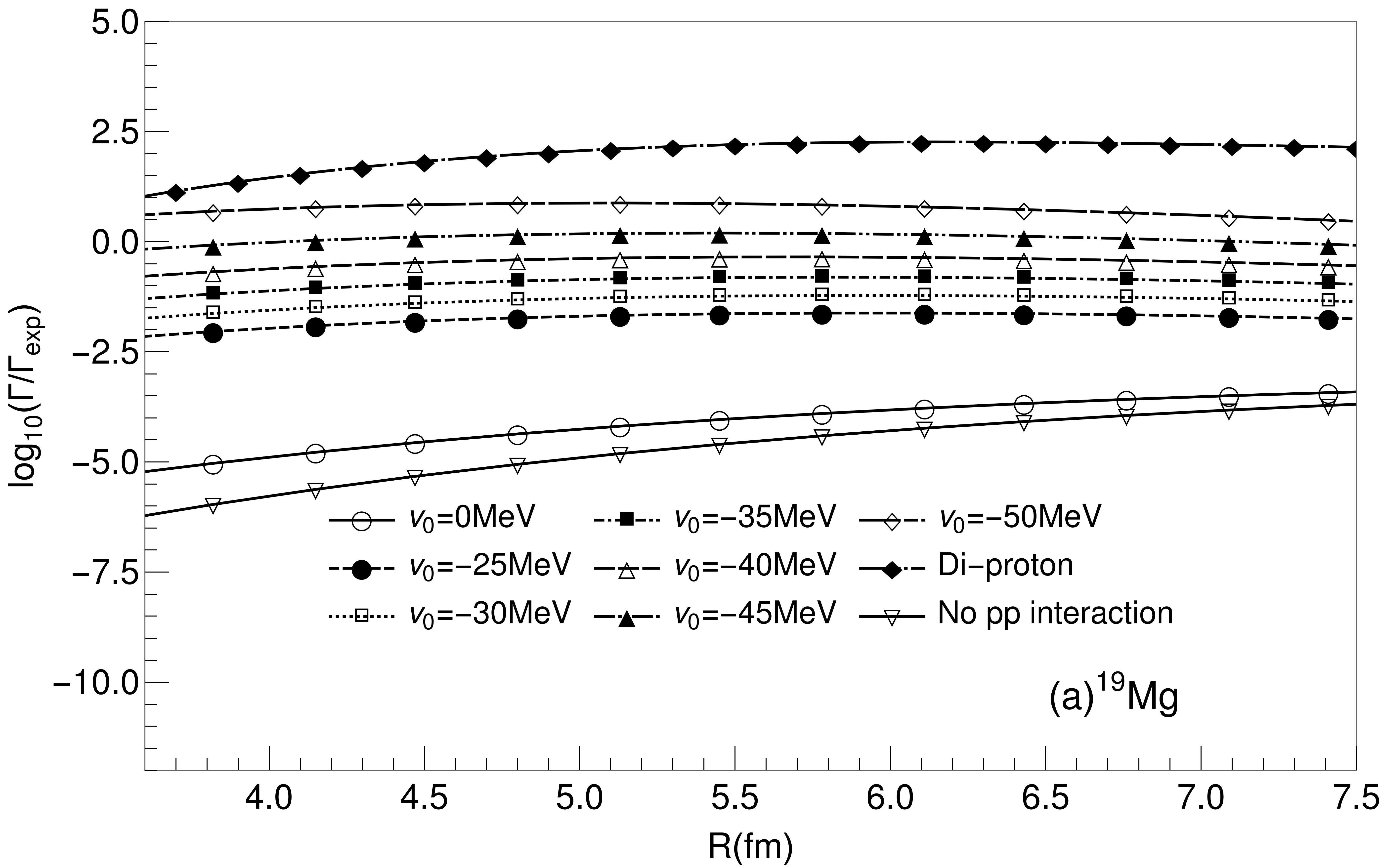}
		\includegraphics[width=\columnwidth]{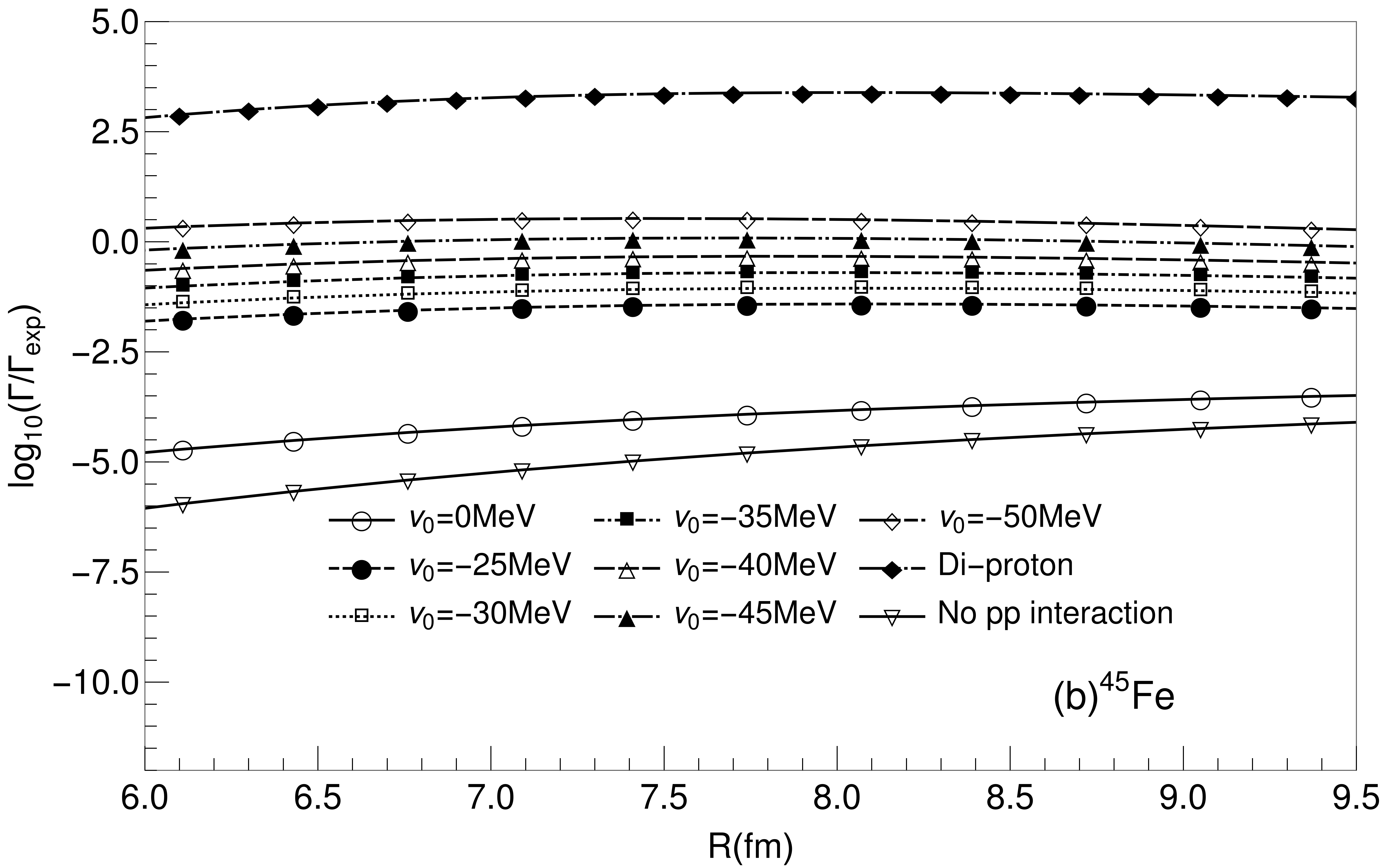}
		\includegraphics[width=\columnwidth]{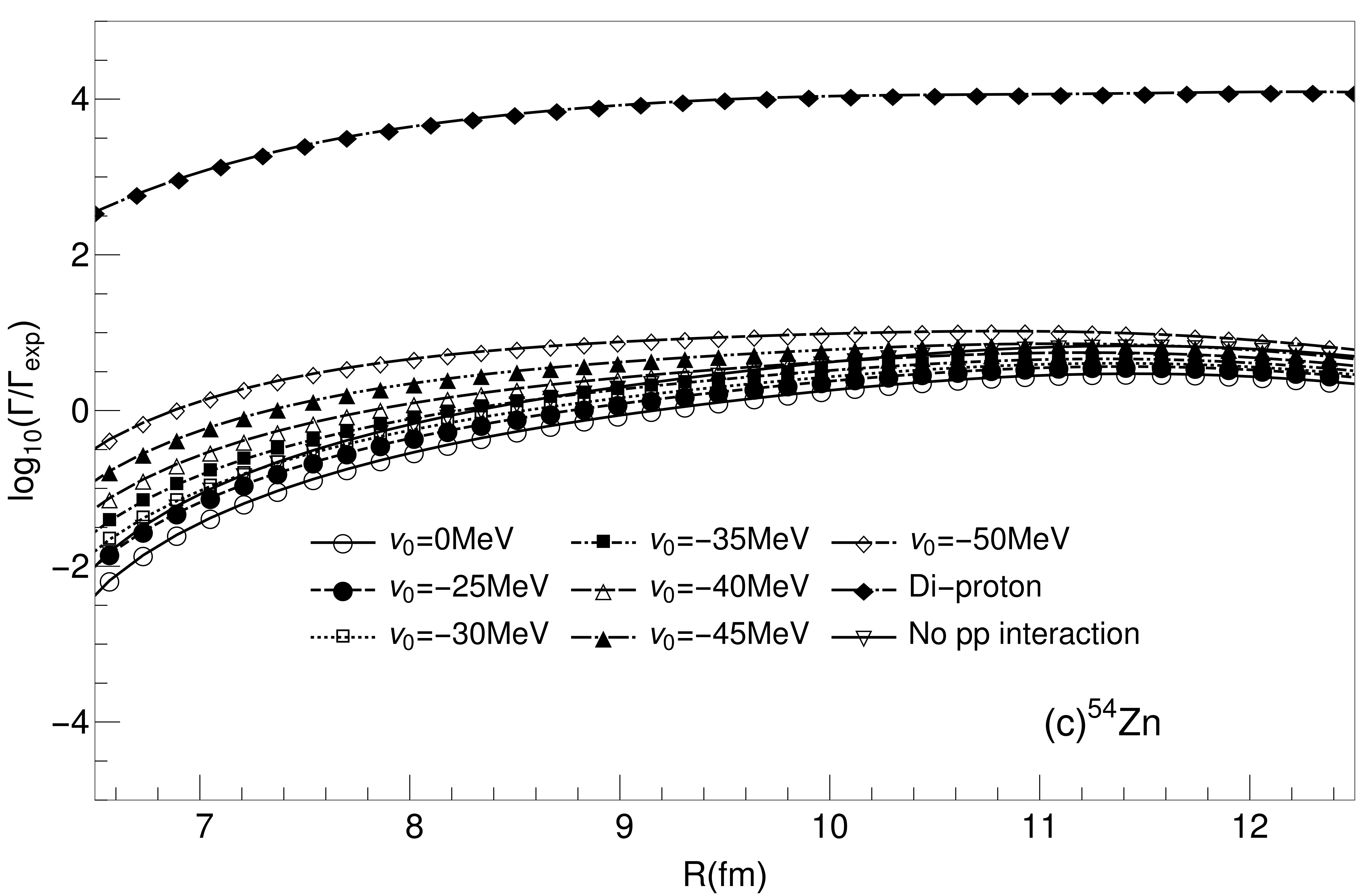}
		\caption{\label{fig:Gamma_vs_V0} Decay width dependence on matching radius, for multiple proton-proton interaction strengths for $^{19}$Mg (a), $^{45}$Fe (b) and $^{54}$Zn (c). The parameter $r_0$ is fixed at $2$~fm in all cases.}
	\end{figure}

	\begin{figure}
		\centering
		\includegraphics[width=\columnwidth]{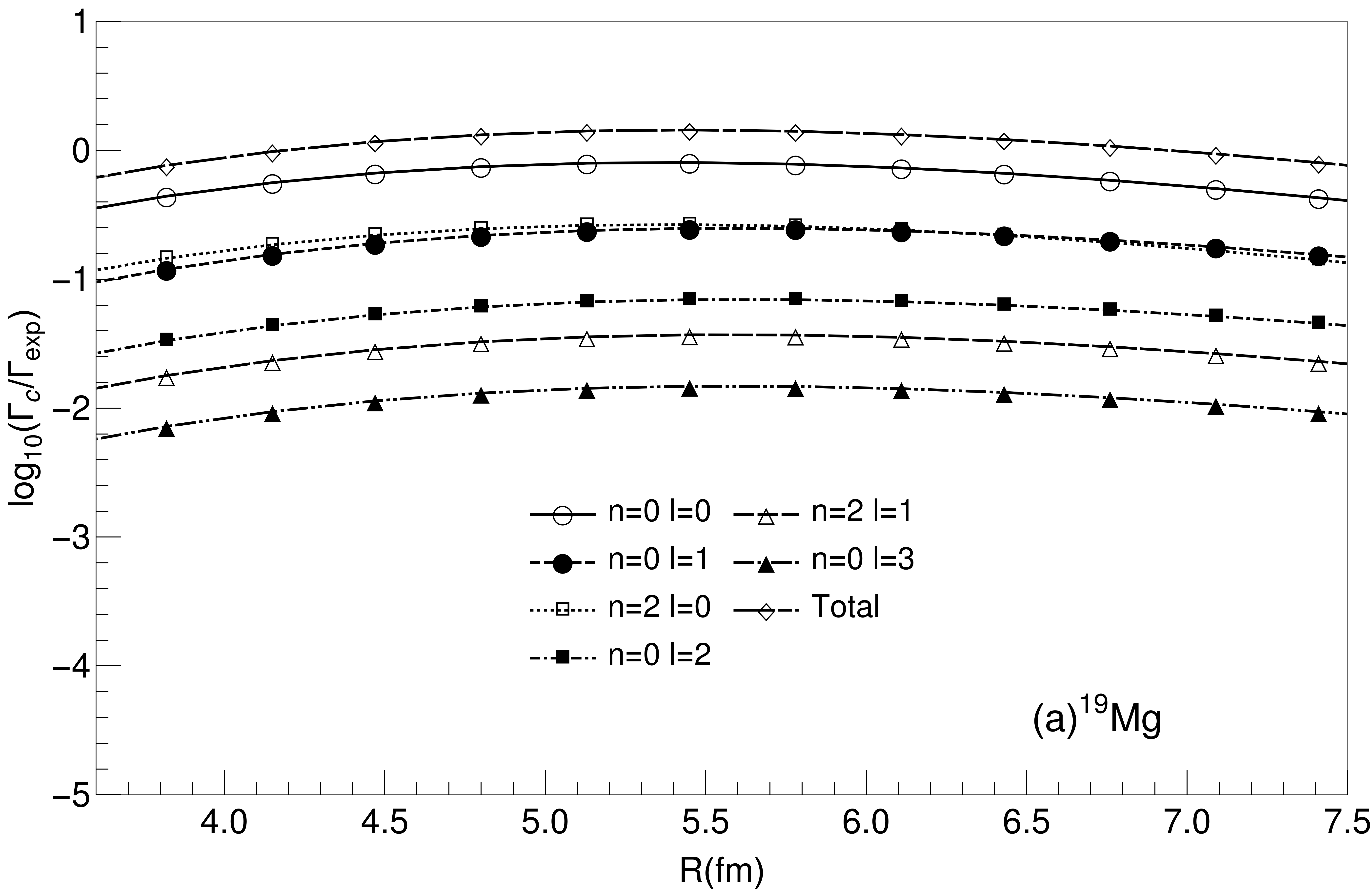}
		\includegraphics[width=\columnwidth]{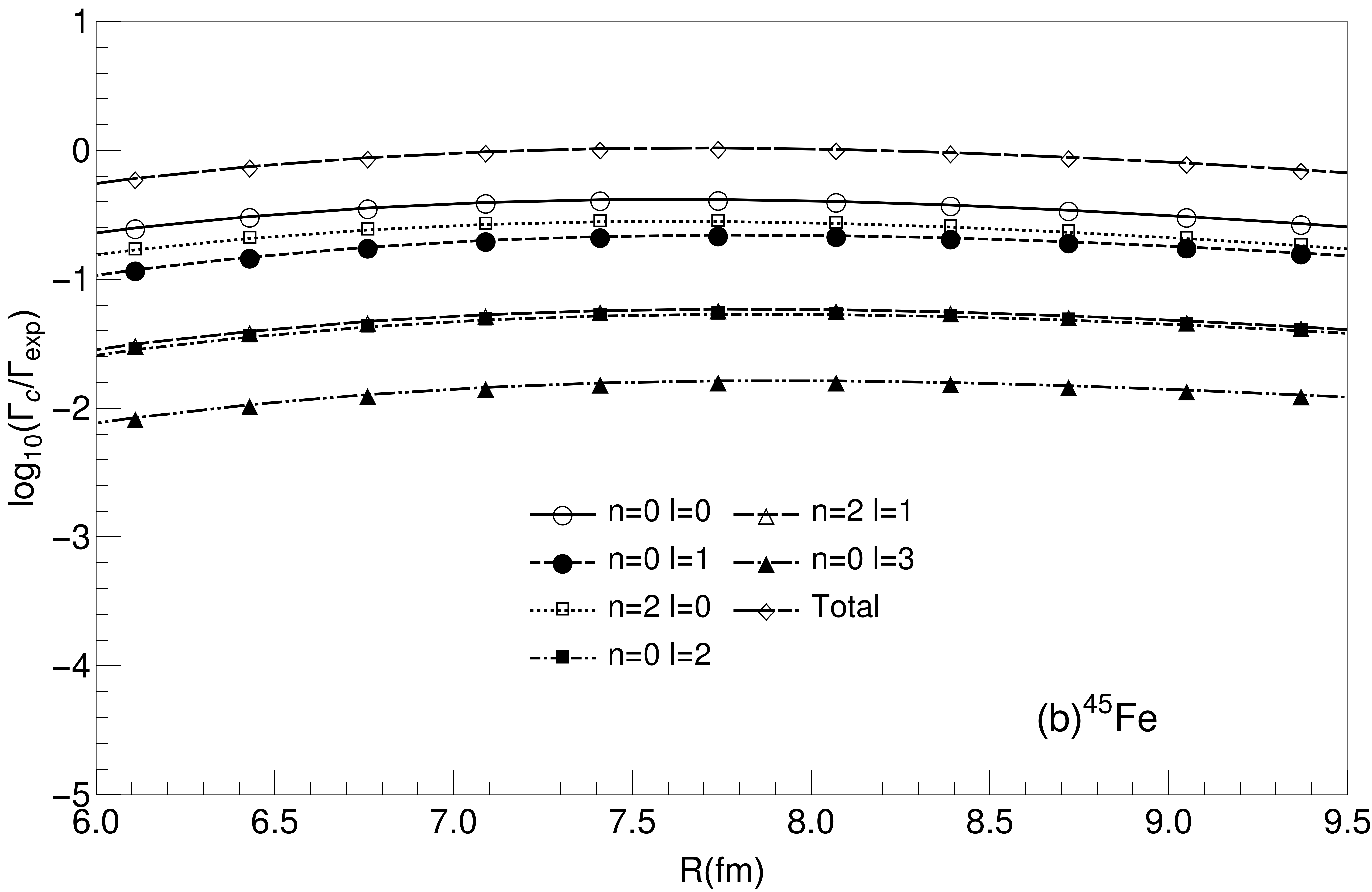}
		\includegraphics[width=\columnwidth]{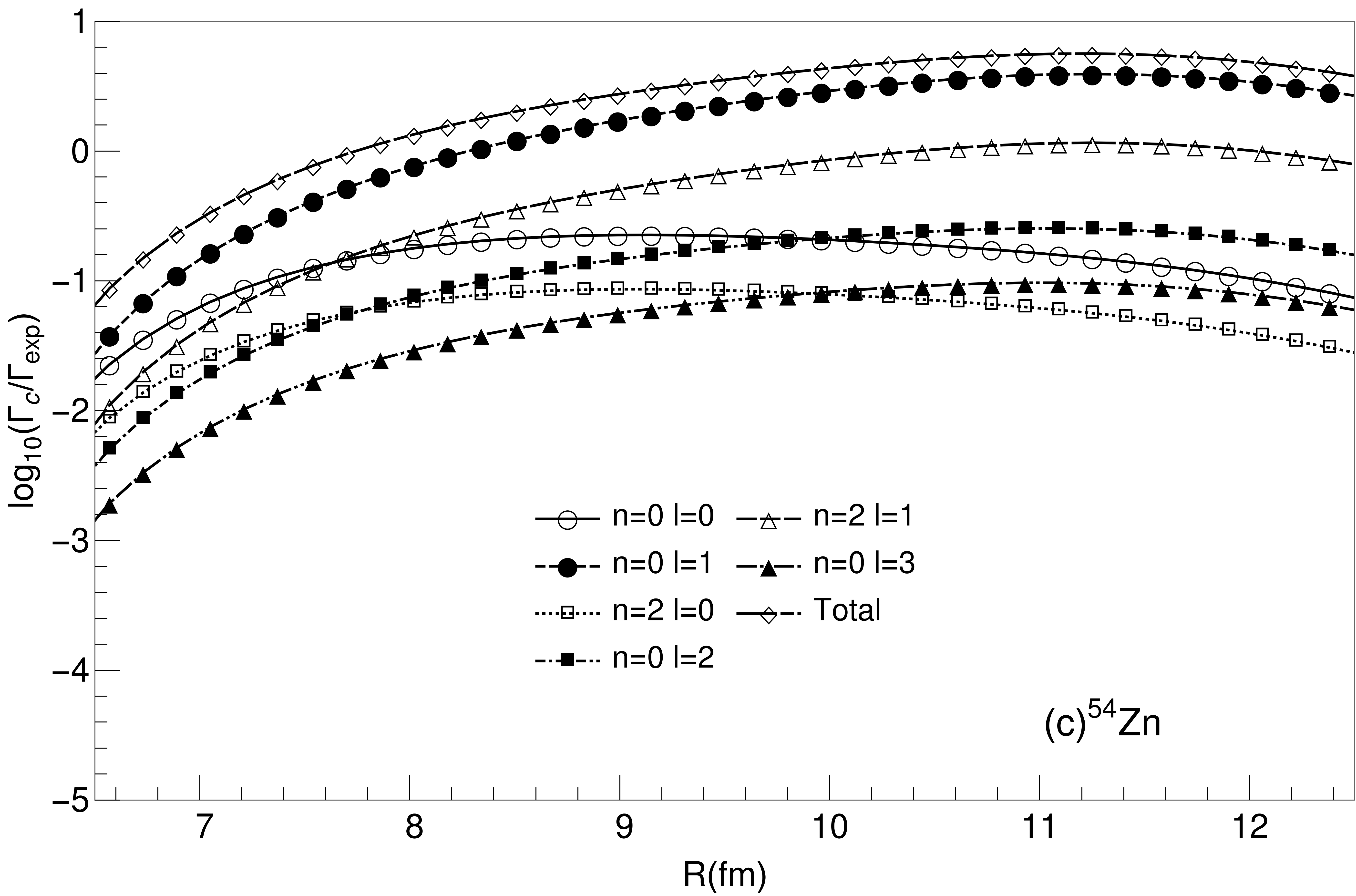}
		\caption{\label{fig:PartailGammas} Partial decay widths of the first 6 dominant channels at $v_0 = -45$ MeV and $r_0=2$~fm as function of the matching radius for $^{19}$Mg (a), $^{45}$Fe (b) and $^{54}$Zn (c). The dashed line with open diamond symbols is the sum of the displayed partial widths.}
	\end{figure}

	We turn now to analyzing the proton pairing effects. Fig. \ref{fig:Gamma_vs_V0} shows the dependence of the decay width as function of the matching radius for $^{19}$Mg, $^{45}$Fe and $^{54}$Zn. We note that the sequential (i.e. no proton interaction) and diproton cases bound the microscopic estimates from below and above respectively. This has been remarked in the past by several authors. It is also important to underline that including only the Coulomb interaction between the protons leads to serious underestimations of the decay width. This is true for both low masses (where the $pN$ and $pp$ potentials are of the same order of magnitude) and at high masses (where the $pN$ interaction is dominant on most of the radial range).
	
	The dependence on the proton pairing strength ($v_0$) shows some remarkable features. First of all, for all three nuclei, the value of $\Gamma$ from our model is in agreement with the experiment for $v_0 \simeq -43 $ MeV, which is a reasonable value, considering the degree of approximation in our work (the "bare" value is $v_0\simeq -35 $ MeV). We also note that the variation of the decay width with $v_0$ decreases as the mass number increases. This effect is caused by 2 components. The first is that the proton pairing strength inside the nucleus decreases with increasing mass. This leads to a smaller gap and, consequently, to a more confined wave function on the surface. The second component is the relative importance of the proton-proton nuclear interaction w.r.t. the proton-nucleus Coulomb interaction. Since the $pN$ Coulomb potential is more than twice stronger than the $pp$ interaction, it is expected that small variations of the latter will not matter much. For the $^{54}$Zn case, the stability plateau is large (between 8 and 11.5fm) at $v_0 = -45$ MeV. Towards the end of this interval, a variation of even 15 MeV in $v_0$ influences the total decay width by a factor less than 5. Notice also that the effective value of the nuclear strength $v_0$ inside nucleus, given by solving BCS equations, is by almost one order of magnitude smaller than its "bare" value in the freee space.
		
	In Fig. \ref{fig:PartailGammas} we plot the partial widths corresponding to the first 6 most important channels, again as function of the matching radius and for the same 3 nuclei. First of all, we note that in all 3 cases there is a dominant channel which is either the monopole (for $^{19}$Mg and $^{45}$Fe) or the $n=0$,$l=1$ channel (for $^{54}$Zn). This is expected due to the monopole centrifugal barrier being lower than on the other channels. For the $^{54}Zn$ case, the internal wavefunction is already built on the $l=1$ channels. The flow from $l=1$ channels towards $l=0$ channels is hindered by the Raynal-Revay coefficients.

	\section{Conclusions}
		
	In this paper we built a semi-microscopic model in hyper-spherical coordinates for the 2 proton emission process. We assumed that the protons are emitted from a paired state and that the transition happens between the ground states of the parent and daughter nuclei. 
	
	By splitting the radial domain in an external region and an internal region, we avoid several difficulties associated to other models in literature. We achieved the plateau condition for decay widths beyond the nuclear radius. More importantly, we show that our model is sensitive to pairing correlations between the emitted protons. We studied the effect of these correlations on 3 nuclei, of significantly different masses. We showed that the partial life-times of these nuclei can be well reproduced using reasonable values for the $pp$ potential ($v_0\approx -45$MeV and $r_0\approx 2$fm). 
	
	As expected, the decay widths predicted with our model lie between the the two extreme mechanisms proposed by Goldansky. In turn, this implies that the 2 proton emission process is a valuable tool for investigating the proton-proton potential both inside the nuclear medium and far away from it.
	 
	\begin{acknowledgments}
	\end{acknowledgments}

	This work was supported by the grant of the Romanian Ministry Education and Research No. PN-18090101/2019-2021.

	\bibliography{bibliography}
\end{document}